# Développement d'instabilités dans un écoulement subsonique se développant au-dessus d'une cavité : mesures synchronisées PIV-LDV

**Thierry M. FAURE**[*,**], **François LUSSEYRAN**[*], **Luc PASTUR**[*,†], **Romain PETHIEU**[*,†], **Philippe DEBESSE**[*,**]

[*]Laboratoire d'Informatique pour la Mécanique et les Sciences de l'Ingénieur, LIMSI-CNRS, B.P. 133, 91403 Orsay Cedex, France
[**]Université Pierre et Marie Curie, Paris VI, 4 place Jussieu, 75252 Paris Cedex 05, France
[†]Université Paris-Sud XI, 91405 Orsay Cedex, France
Email auteur correspondant : thierry.faure@limsi.fr

L'interaction entre une couche limite et une cavité est une configuration aérodynamique de référence qui se rencontre dans de nombreuses applications environnementales, aéronautiques, automobiles et industrielles. La couche de cisaillement qui se développe est caractérisée par l'excitation d'un ou de plusieurs modes. Les conditions pour lesquelles cet écoulement présente au moins deux modes spectraux correspondent à une forme de compétition, en particulier à grandes vitesses [1]. L'intermittence de mode qui en résulte est aussi présente lorsque la vitesse extérieure et faible [2]. Cependant aucun mécanisme n'a encore été démontré, ni aux grandes ni aux basses vitesses. Cette étude a pour objectif d'établir le lien entre l'alternance modale et la structure spatiale de l'écoulement. Une couche limite laminaire interagit avec une cavité ouverte pour deux nombres de Reynolds moyens. Le développement de la couche de cisaillement induit la création de structures cohérentes à l'intérieur de la cavité mises en évidence par des visualisations dans différents plans. Des mesures synchronisées par vélocimétrie par images de particules (PIV) et vélocimétrie laser Doppler (LDV) sont réalisées pour disposer à la fois de champs spatiaux et d'une mesure ponctuelle résolue en temps. Il est alors possible d'effectuer un traitement des champs PIV en phase à partir d'une décomposition aux valeurs singulières du signal LDV.

## 1 Dispositif expérimental

L'écoulement est crée par un ventilateur centrifuge placé en amont de la chambre de tranquillisation (Figure 1-a). L'injection de marqueur s'effectue en entrée du ventilateur. Il s'agit soit de fumée basse densité pour les visualisations de l'écoulement, soit de particules de phtalate de di-n-octyle pour la vélocimétrie, d'un diamètre moyen de 1 $\mu$m. Un conduit se terminant par du nid d'abeille et un convergent amène l'écoulement vers la section d'essais, constituée d'une plaque plane munie d'un bord d'attaque elliptique afin de fixer l'origine de la couche limite. La longueur de la plaque A = 300 mm permet de fournir une couche limite laminaire établie. Les réflexions lumineuses dues aux parois sont minimisées par l'utilisation d'un verre antireflet de 2 mm d'épaisseur pour l'ensemble de la section d'essais. La hauteur de la cavité est H = 50 mm et son envergure S = 300 mm (Figure 1-b), les extrémités de la cavité selon cette direction étant les parois verticales de la soufflerie. Le rapport de forme de la cavité R = L / H est égal à 2. En sortie de la soufflerie, l'air est rejeté dans la salle de mesure. La vitesse extérieure $U_e$ est mesurée par LDV 102 mm en amont de la cavité et 25,5 mm au-dessus de la plaque plane. Ce point de mesure est situé dans l'écoulement extérieur suffisamment en amont de la cavité pour ne subir aucune perturbation de l'instabilité qui se développe au-dessus de la cavité. Deux nombres de Reynolds, déterminés avec la longueur L et la vitesse $U_e$, sont étudiés, respectivement 8470 et 13900, ce qui correspond à des vitesses extérieures respectives de 1,27 m.s$^{-1}$ et 2,09 m.s$^{-1}$. L'origine du système de coordonnées est prise au bord amont de la cavité à mi-envergure, l'axe x est dans la direction de l'écoulement, l'axe y normal à la plaque amont et l'axe z selon l'envergure. La paroi supérieure de la section d'essais, située à D = 75 mm au-dessus de la cavité, n'a aucune influence sur la couche de cisaillement. L'épaisseur de la couche limite sur cette paroi est inférieure à 10 mm et n'a pas d'influence sur l'écoulement extérieur.

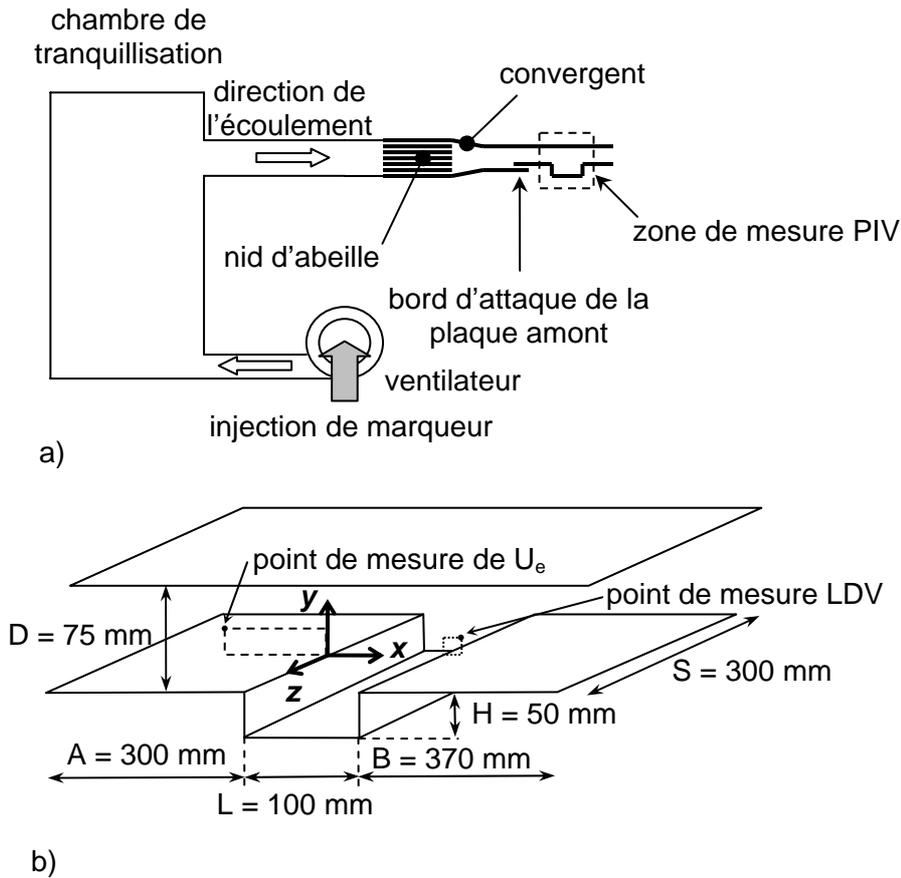

*Figure 1 : Dispositif expérimental : a) soufflerie basse vitesse, b) section d'essais et système de coordonnées*

## 2   Visualisations de l'écoulement

Les deux longueurs d'ondes bleue et verte d'un laser continu argon-ion permettent de générer des plans lumineux pour visualiser l'écoulement selon différentes directions [3]. En particulier, deux caméras de fréquence d'acquisition 20 Hz synchronisées et munies chacune d'un filtre permettent de visualiser l'écoulement dans deux plans verticaux parallèles (Figure 2). La Figure 3 présente une visualisation de l'écoulement pour $U_e = 1,27$ m.s$^{-1}$ qui met en évidence la présence d'un tourbillon principal dans la partie aval de la cavité et d'un tourbillon secondaire contrarotatif dans la partie amont, l'ensemble étant fortement instationnaire. L'injection de fumée au bord aval reste la même entre ces deux plans, ce qui montre que l'injection de fluide dans la cavité est plutôt bidimensionnelle, tandis que le reste de l'écoulement présente des motifs différents sur 30 mm. Une visualisation dans un plan horizontal confirme cette observation (Figure 4) avec l'injection au bord aval quasi-uniforme selon l'envergure, tandis que des structures tridimensionnelles se développent dans le reste de la cavité. En particulier, les structures cohérentes présentes du côté du bord amont ont pu être identifiées comme la déstabilisation de tourbillons de Görtler [4]. L'écoulement à l'intérieur de la cavité est semblable à ce que l'on observe dans une cavité fermée par un couvercle entrainé [5, 6]. Il n'est plus possible d'identifier de structure cohérente dans l'écoulement pour $U_e = 2,09$ m.s$^{-1}$ car la fumée se dissipe rapidement sous l'effet de la turbulence qui se développe à l'intérieur de la cavité.

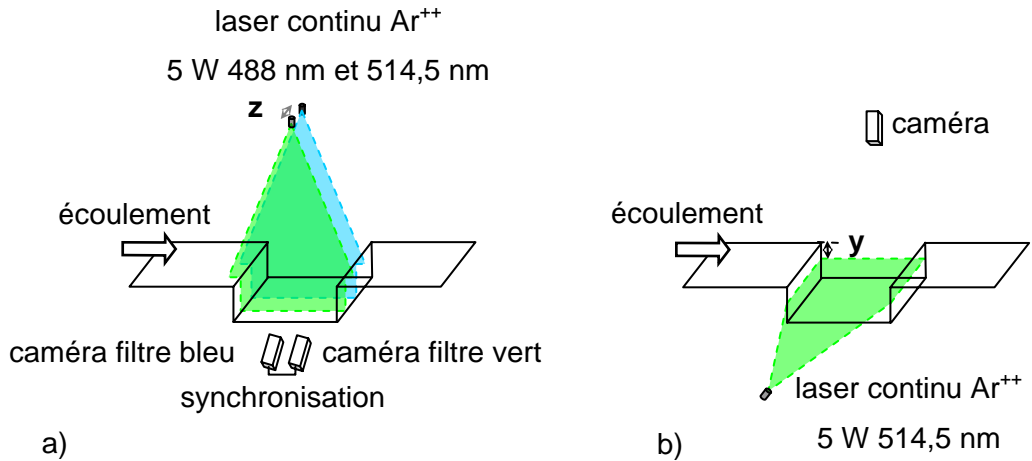

*Figure 2 : Montage pour la visualisation a) dans deux plans verticaux b) dans un plan horizontal*

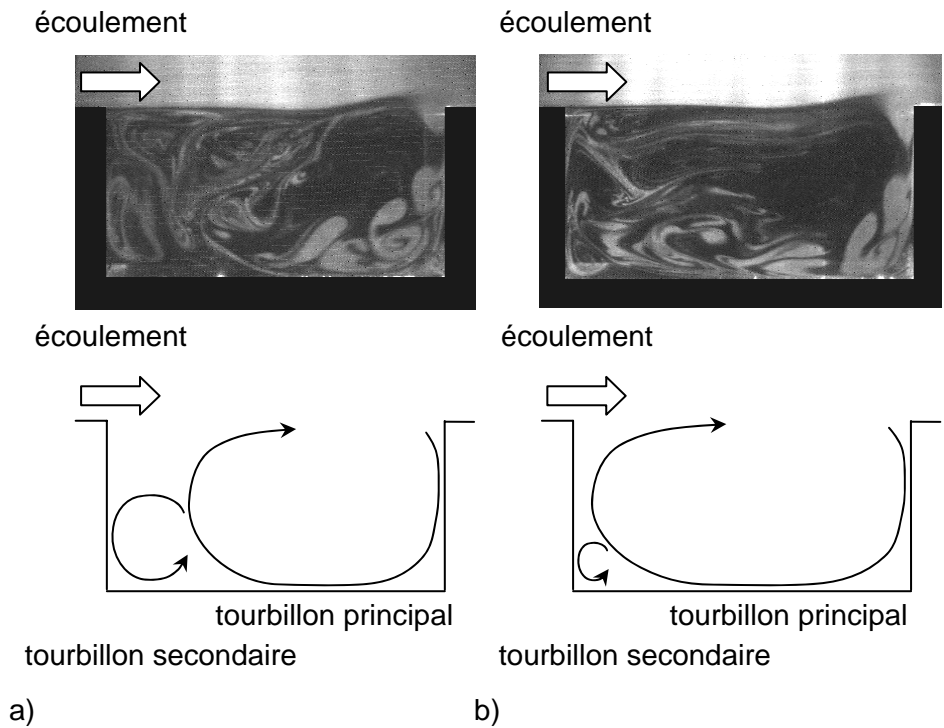

*Figure 3 : Visualisations synchronisées dans deux plans parallèles et schémas de l'écoulement pour $U_e = 1,27$ m.s$^{-1}$ : a) z = 0 mm, b) z = 30 mm*

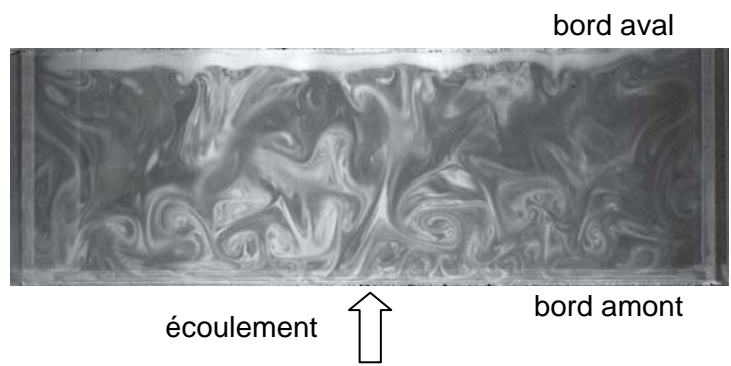

*Figure 4 : Visualisation dans un plan horizontal situé à y = -15 mm pour $U_e = 1,27$ m.s$^{-1}$*

## 3   Mesures synchronisées PIV-LDV

Des mesures synchronisées entre la PIV et la LDV sont effectuées afin d'obtenir à la fois des champs spatiaux de vitesse et des signaux résolus en temps en un point particulier situé en aval de la cavité (Figure 5). Le système PIV comprend un laser YAG pulsé fournissant une nappe laser de longueur d'onde 532 nm d'épaisseur maximum 0,25 mm et une caméra 20 Hz pour l'enregistrement des images. Le système LDV utilise la longueur d'onde bleue (488 nm) du laser continu Argon-ion et fonctionne en mode de diffusion avant pour acquérir la composante axiale de vitesse. Le volume de mesure est placé 15 mm en aval de la cavité et 15 mm au-dessus de la plaque aval. Afin de s'affranchir de la longueur d'onde 488 nm sur les images PIV, un filtre passe-bande centré sur la longueur d'onde du YAG et d'une largeur de bande de 10 nm est monté sur l'objectif de la caméra.

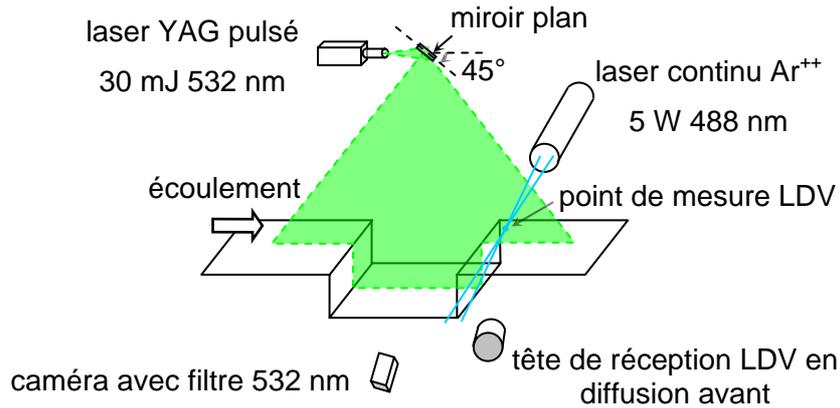

*Figure 5 : Montage de mesures synchronisées PIV-LDV*

## 4   Traitement statistique des champs de vitesse PIV

Les images sont d'abord traitées par soustraction d'une image de bruit de fond prise sans écoulement [7]. Les vitesses PIV sont obtenues par un algorithme de flot optique utilisant une programmation dynamique orthogonale [8]. La résolution des mesures est de 1 / 32$^e$ de pixel, ce qui correspond pour $U_e$ = 1,27 m.s$^{-1}$ à une incertitude de vitesse relative de 0,15 %. Le champ moyen de vitesse est déterminé à partir de 1100 champs instantanés (Figure 6).

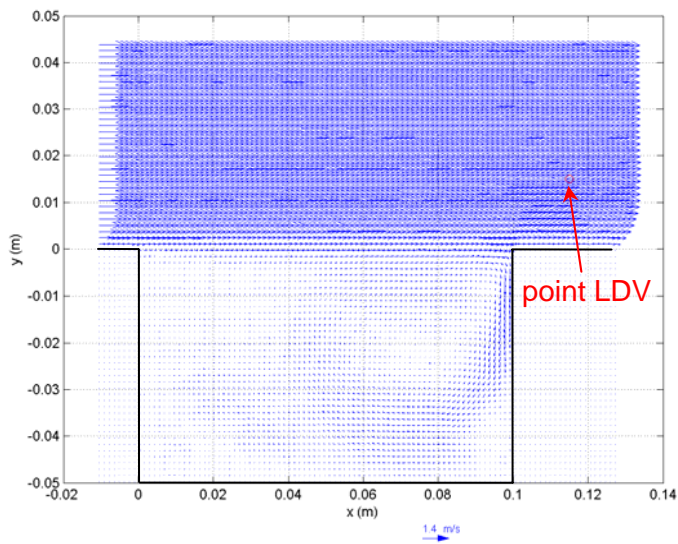

*Figure 6 : Champ PIV moyen pour $U_e$ = 1,27 m.s$^{-1}$ (étant donnée la forte résolution de la méthode PIV employée qui est d'un vecteur par pixel et pour que les vitesses demeurent visibles, seul un vecteur sur dix est représenté)*

## 5   Analyse modale de la vitesse LDV

L'analyse par spectrogramme du signal LDV $U_x(t)$ met en évidence un mode d'excitation caractéristique de la couche de cisaillement, pour $U_e$ = 1,27 m.s$^{-1}$, à la fréquence $f_1$ = 12,8 Hz qui demeure toujours présent dans le signal temporel (Figure 7). Pour $U_e$ = 2,09 m.s$^{-1}$ on observe deux modes d'oscillation intermittents, aux fréquences $f_1$ = 23 Hz et $f_2$ = 31 Hz avec une succession de passages entre ceux-ci (Figure 8).

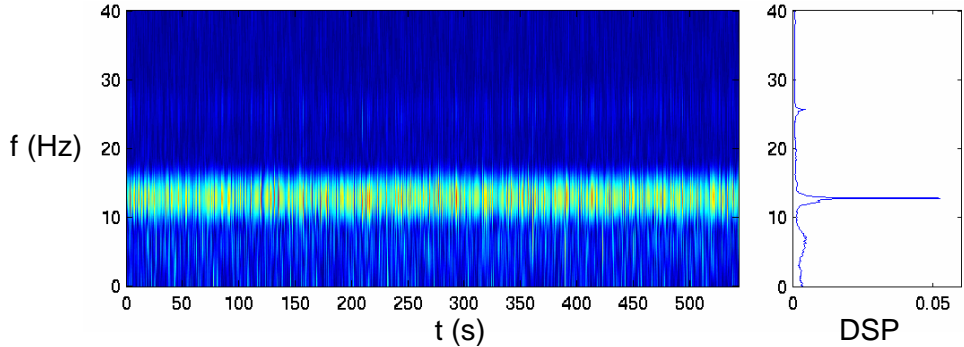

*Figure 7 : Spectrogramme (à gauche) et densité spectrale de puissance (à droite) de la vitesse $U_x$ pour $U_e$ = 1,27 m.s$^{-1}$*

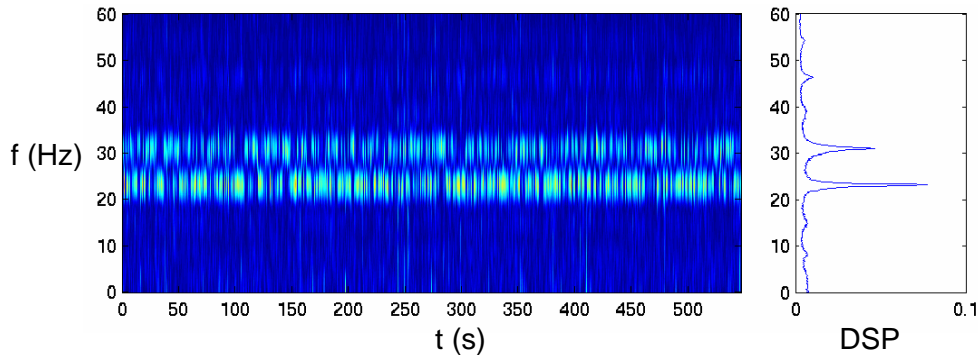

*Figure 8 : Spectrogramme (à gauche) et densité spectrale de puissance (à droite) de la vitesse $U_x$ pour $U_e$ = 2,09 m.s$^{-1}$*

## 6   Traitement des champs PIV en phase par rapport à la vitesse LDV

L'acquisition synchronisée entre la PIV et la LDV fournit la composante de vitesse axiale mesurée en un point à une haute fréquence d'acquisition (1,5 kHz) et des champs spatiaux de deux composantes de vitesse sous-échantillonnés par rapport à la LDV (10 Hz). On effectue alors un traitement des champs PIV par moyenne de phase, à partir du portrait de phase du signal LDV obtenu par une décomposition en valeurs singulières [9]. Pour $U_e$ = 1,27 m.s$^{-1}$, le signal LDV $U_x(t)$ est tout d'abord ré-échantillonné à une fréquence multiple de la fréquence des champs PIV. Il est ensuite filtré au moyen d'un filtre passe-bande de largeur de bande 1 Hz autour de la fréquence correspondant au mode caractéristique de la couche de cisaillement, et sert à la construction de la matrice des retards B. Une décomposition aux valeurs singulières est alors réalisée :

$$B = U \cdot D \cdot V^T$$

où D est la matrice des valeurs singulières qui sert à la construction de la matrice :

$$X = U \cdot D = B \cdot V$$

Les deux premières colonnes de la matrice X permettent alors la réalisation du portrait de phase ($X_1$, $X_2$), chaque point étant repéré par son rayon r et sa phase $\varphi$ (Figure 9). Les champs PIV correspondants sont alors moyennés par secteurs en 16 phases de résolution angulaire 22,5° (Figure 10). La longueur d'onde des instabilités est déterminée en ajustant le profil de la composante de vitesse $U_y$ le long de l'axe y = 0 sur la fonction d'une onde spatiale amplifiée :

$$U_y = A + Be^{\beta x} \cos\left(\frac{2\pi}{\lambda}x + \varphi\right)$$

On obtient ainsi $\lambda = 0,046$ m. La vitesse de convection des instabilités d'une couche de cisaillement créée par deux écoulements de vitesses $U_1$ et $U_2$ est :

$$U_c = \frac{U_1 + U_2}{2}$$

Dans le cas présent et en première approximation $U_1 \approx U_e$ et $U_2 \approx 0$ ce qui donne :

$$U_c \approx \frac{U_e}{2} = 0,635 \text{ m.s}^{-1}$$

Cette valeur est proche (à moins de 8%) de celle obtenue par ajustement sur $U_y$ :

$$U_c = \lambda f = 0,588 \text{ m.s}^{-1}$$

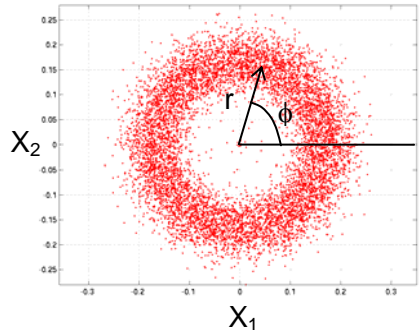

*Figure 9 : Portrait de phase du signal LDV pour $U_e = 1,27$ m.s$^{-1}$*

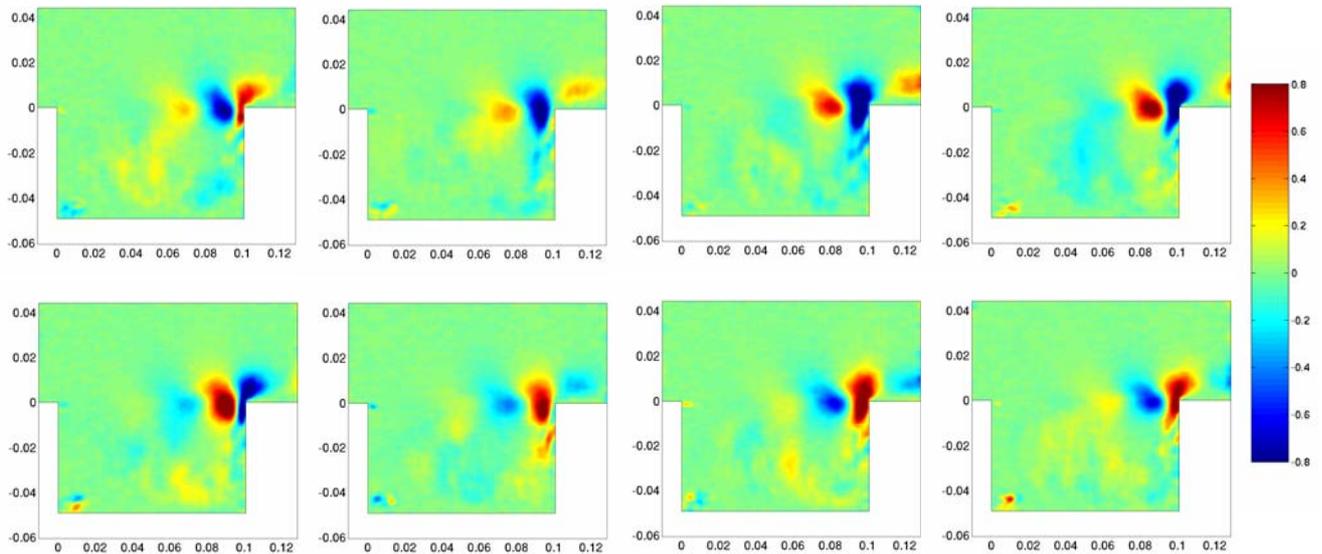

*Figure 10 : Composante $U_y$ de vitesse PIV en phase avec le signal LDV pour $U_e = 1,27$ m.s$^{-1}$*

Le même traitement est réalisé pour $U_e = 2,09$ m.s$^{-1}$ où deux modes sont présents. Cette fois-ci, le filtrage est effectué successivement autour du premier ou du second mode, et le portrait de phase LDV fait apparaître une forte densité de points autour de deux rayons caractéristiques de chaque mode. La discrimination d'un mode par rapport au second s'effectue en ne prenant que les champs PIV correspondant à un rayon supérieur au rayon du minimum de densité (Figure 11). Les champs moyens en phase de la composante de vitesse $U_y$ sont présentés sur la Figure 12 pour un filtrage sur le premier mode et sur la Figure 13 pour un filtrage sur le second mode. L'ajustement d'un profil de vitesse sur une onde spatiale amplifiée fournit ainsi pour le premier mode $\lambda_1 = 0,047$ m ($f_1 = 23,2$ Hz) et une vitesse $U_{c,1} = 1,09$ m.s$^{-1}$ et pour le second mode $\lambda_2 = 0{,}0378$ m ($f_2 = 31{,}0$ Hz) et une vitesse $U_{c,2} = 1{,}17$ m.s$^{-1}$, vitesses à rapprocher de la vitesse de convection théorique $U_c = 1{,}045$ m.s$^{-1}$.

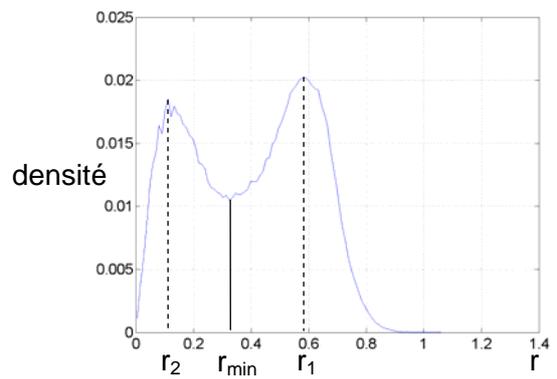

*Figure 11 : Distribution des points du portrait de phase en fonction de leur rayon r pour un filtrage sur le premier mode pour $U_e = 2{,}09$ m.s$^{-1}$*

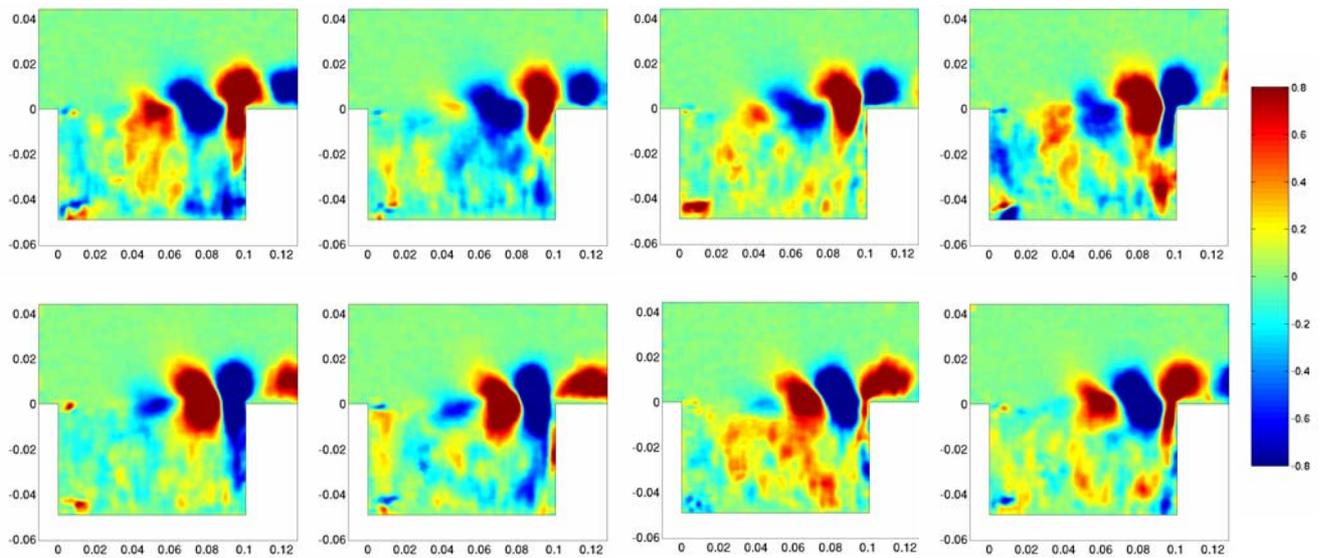

*Figure 12 : Composante $U_y$ de vitesse PIV en phase avec le premier mode pour $U_e = 2{,}09$ m.s$^{-1}$*

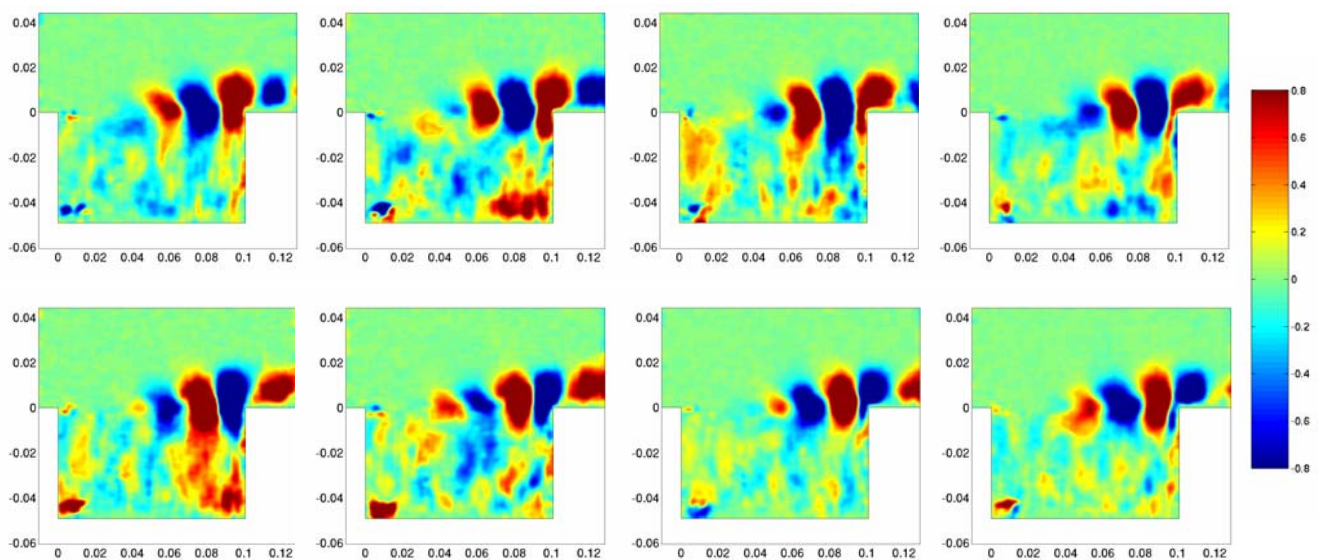

*Figure 13 : Composante $U_y$ de vitesse PIV en phase avec le second mode pour $U_e = 2{,}09$ m.s$^{-1}$*

## 7 Conclusion

L'interaction entre une couche limite laminaire et une cavité de rapport de forme R = 2, pour des nombres de Reynolds de l'ordre de 10000, engendre un écoulement de cavité fortement tridimensionnel à l'exception du développement de la couche de cisaillement et de l'injection de fluide dans la cavité qui restent essentiellement bidimensionnels. On observe un mode caractéristique de cisaillement pour $U_e$ = 1,27 m.s$^{-1}$ et deux modes intermittents pour $U_e$ = 2,09 m.s$^{-1}$. L'acquisition synchronisée PIV-LDV permet de traiter les champs PIV par moyenne de phase à partir du portrait de phase du signal LDV obtenu par une décomposition aux valeurs singulières. Cette analyse en phase est effectuée dans le cas de la présence d'un seul ou de deux modes. Il est ainsi possible de déterminer les longueurs d'onde caractéristiques associées à chacun des deux modes d'oscillation de la couche de cisaillement, longueurs d'ondes très difficiles à estimer sur les champs instantanés de vitesse.

## 8 Références